\documentclass[12pt]{iopart}

\usepackage{iopams}

\newtheorem{theorem}{Theorem}[section]

\newtheorem{proposition}[theorem]{Proposition}
\newtheorem{corollary}[theorem]{Corollary}

\newtheorem{example}[theorem]{Example}

\def\qed{\hfill $\Box$\medskip}

\def\IC{{\bf C}}

\def\bu{{\bf u}}

\def\bx{{\bf x}}
\def\by{{\bf y}}

\def\diag{{\rm diag}\,}
\def\tr{{\rm tr}\,}
\def\la{{\langle}}
\def\ra{{\rangle}}

\def\dfrac{\displaystyle\frac}
\def\[{\left [}
\def\]{\right ]}
\def\({\left (}
\def\){\right )}
\def\la{{\langle}}
\def\ra{{\rangle}}
\def\Lra{{\ \Leftrightarrow\ }}
\def\Ra{{\ \Rightarrow\ }}

\begin{document}
\openup .98 \jot

\title[Criteria and new classes of k-positive maps]
{Criteria and new classes of k-positive maps}

\author{Jinchuan Hou$^1$, Chi-Kwong Li$^2$, Yiu-Tung Poon$^3$, Xiaofei Qi$^4$, and Nung-Sing Sze$^5$}

\address{$^1$ Faculty of Mathematics, Taiyuan University of Technology, Taiyuan, 030024, China}

\address{$^2$ 100 Talent Scholar, Faculty of Mathematics, Taiyuan University of Technology, \\
\hspace{2mm} Taiyuan, 030024, China; Department of Mathematics, College of William and Mary, \\
\hspace{2mm} Williamsburg, VA 23187, USA}

\address{$^3$ Department of Mathematics, Iowa State University, Ames, IA 50011, USA}

\address{$^4$ Department of Mathematics, Shanxi University, Taiyuan, 030006, China}

\address{$^5$ Department of Applied Mathematics, Hong Kong Polytechnic University, \\
\hspace{2mm} Hung Hom, Hong Kong}

\ead{
\hspace*{-6mm} jinchuanhou@yahoo.com.cn, 
ckli@math.wm.edu,\\
\hspace*{11mm} ytpoon@iastate.edu,
qixf1980@126.com,
raymond.sze@polyu.edu.hk}

\begin{abstract}
We study $k$-positive maps on operators. Proofs are given to
different positivity criteria. Special attention is on
positive maps arising in the study of quantum information science.
Results of other researchers are extended and improved. New classes
of positive maps are constructed. Some open questions are answered.
\end{abstract}

\maketitle

\section{Introduction}

Denote by $B(H,K)$ the set of bounded linear operators from the
Hilbert space $H$ to the Hilbert space $K$, and write $B(H,K) =
B(H)$ if $H = K$. Let $B(H)^+$ be the set of positive semidefinite
operators in $B(H)$. If $H$ and $K$ have dimensions $n$ and $m$
respectively, we identify $B(H,K)$ with the set $M_{m,n}$ of
$m\times n$ matrices, and write $M_{n,n} = M_n$, and $B(H)^+ =
M^+_n$.

A linear map $L: B(H) \rightarrow B(K)$ is positive if $L(B(H)^+)
\subseteq B(K)^+$. For a positive integer $k$, the map $L$ is
$k$-positive if the map $I_k \otimes L: M_{k}(B(H)) \rightarrow
M_{k}(B(K))$  is positive, where $(I_k \otimes L)(A) = (L(A_{ij}))$
for any $A = (A_{ij})_{1 \le i,j\le k}$ with $A_{ij} \in B(H)$. A
map is completely positive if it is $k$-positive for every positive
integer $k$. The study of positive maps has been the central theme
for many pure and applied topics; for example, see
\cite{Choi,H2,K,LP,MN}. In particular, the study has attracted a lot
of attention of physicists working in quantum information science in
recent decades, because positive linear maps can be used to
distinguish entanglement of  quantum states (see \cite{Hor}). There
is considerable interest in finding positive maps that are not
completely positive, which can be applied to detect entangled states
(see, for example, \cite{AU,CJW,CKL,CK,CK2,CK3, CK4, EK, HHH1,H,
HLPS, LMM, QH, SA, SA2} and the references therein). Completely positive
linear maps have been studied extensively by researchers. However, the
structure of positive linear maps is still unclear even for the
finite dimensional case (\cite{C2,De, H2,P}).

In this paper, we give a brief summary of some frequently used
criteria of $k$-positive maps on operators for convenient
reference. New proofs are given to these different positivity
criteria. Special attention is on positive maps arising in the
study of quantum information science. Furthermore, some of the
existing results are extended and improved and some open problems
are answered.

The paper is organized as follows. Sections 2 and 3 summarize some
basic known criteria for the different types of $k$-positive maps
and several new criteria for elementary operators by using
$k$-numerical range of operators are presented (Propositions 2.1-2.2
and 3.1-3.2). In Section 4, we refine the results of
Chru\'{s}ci\'{n}ski and Kossakowski in \cite{CK} (see Propositions
\ref{4.2} - \ref{4.3}) by the tools introduced in Section 3. In
Section 5, we discuss a family of positive maps, called $D$-type
positive maps, which is a generalization of Choi's maps and was
often used in quantum information theory. We give a necessary and
sufficient condition for such maps to be $k$-positive (Proposition
5.1 and Corollary 5.2). Section 6 is devoted to illustrate the
application of results in Section 5 to   constructing   new positive
$D$-type linear maps (Examples \ref{6.1}, \ref{6.6} and \ref{6.7},
Propositions \ref{6.2} and \ref{6.3}). In Section 7, we consider the
decomposability of positive linear maps, propose a new class of
decomposable positive maps, and answer an open problem (Proposition
7.2). Section 8 is a short conclusion.

\section{Basic criteria}

In this section, we reviewed several equivalent conditions of
$k$-positivity that will be used in the subsequent discussion.
Some of these conditions are known; see for example \cite{Choi,EK, LMM, SSZ}. 
Self-contained proofs are given here for completeness.
In the following,  a vector of $H$ will be
denoted by $|x\ra$ and $\la x|$ is defined to be the dual vector of
the vector $|x\ra$ in the dual space of $H$.

\begin{proposition} \label{2.1}
Suppose $L: B(H) \rightarrow B(K)$ is a linear map continuous under strong operator topology.
The following are equivalent.
\begin{enumerate}
\item[{\rm (a)}] $L$ is $k$-positive, i.e., $I_k \otimes L$ is positive.

\item[{\rm (b)}] $(I_k\otimes L)(P)$ is positive semi-definite for any rank
one orthogonal projection $P \in M_k(B(H))$.

\item[{\rm (c)}] For any (orthonormal) subset $X=\{|x_1\ra, \dots, |x_k\ra\} \subseteq H$,
the operator matrix defined by $L_X=(\, L(|x_i\ra \la x_j|)\, )_{1\le i,j\le k}$ is positive semi-definite.
\end{enumerate}
\end{proposition}

\it Proof. \rm The implications (a) $\iff$ (b) $\Longrightarrow$ (c) are
clear because the set of finite rank positive operator is strongly
dense in $B(H)^+$ and $L$ is strongly continuous.  To prove (c)
$\Rightarrow$ (b), one only needs to check the condition for
orthonormal set $\{|x_1\ra, \dots, |x_k\ra\} \subseteq H$.
For every $|z\ra \in H^{\oplus k}$,
write $|z\ra = \sum_{i=1}^k |e_i \ra \otimes|z_i\ra$
where $|z_i \ra \in H$ and $\{|e_i\ra\}_{i=1}^k$ is the canonical basis of $\IC^k$,
and define the finite rank operator $Z=\sum_{i=1}^k|z_i\ra\la e_i|$.
Consider the singular value decomposition (a.k.a. the Schmidt
decomposition in the context of quantum information science) of $Z = \sum_{i=1}^k |y_i\ra \la x_i|$,
one can get a decomposition $|z\ra = \sum_{j=1}^k  |y_j\ra \otimes  |x_j\ra $,
where $\{|y_1\ra, \dots, |y_k\ra\}$ is an orthogonal set in $\IC^k$,
and $\{|x_1\ra, \dots, |x_k\ra\}$ is an orthonormal set in $H$. Let
$Y=\(\sum_{i=1}^k|y_i\ra\la e_i|\)\otimes I_H$. Then
$$(I_k\otimes L)( |z\ra  \la z|) =
(I_k\otimes L)\( \( \sum_{j=1}^k |y_j\ra |  x_j\ra \) \( \sum_{j=1}^k \la x_j |\la y_j| \)\)
=YL_X Y^{\dag}$$
is positive semi-definite  by assumption.
\qed

Suppose $L: M_n \rightarrow B(K)$ is a linear map. Let $\{E_{11}, E_{12}, \dots, E_{nn}\}$
be the standard basis for $M_n$. The Choi matrix $C(L)$ is the operator matrix
with $(L(E_{ij}))_{1 \le i, j \le n}$. Clearly, there is a one-one correspondence
between a linear map $L$ and the Choi matrix $C(L)$. One can use the Choi matrix to determine whether the
map $L$ is $k$-positive.

\begin{proposition} \label{2.2} Let $L: M_n \rightarrow B(K)$ and $1 \le k \le n$.
The following are equivalent.
\begin{enumerate}

\item[{\rm (a)}] $L$ is $k$-positive.

\item[{\rm (b)}] $\la x|C(L)|x\ra \geq 0$ for all $|x\ra = \sum_{p=1}^k
|y_p\ra \otimes |z_p\ra$ with $|y_p\ra \otimes |z_p\ra \in \IC^n\otimes K$.

\item[{\rm (c)}] $(I_n\otimes P)C(L)(I_n\otimes P)$ is positive semi-definite
for any rank-$k$ orthogonal projection $P \in B(K)$.
\end{enumerate}
\end{proposition}

\it Proof. \rm (a) $\Lra$ (b) : First consider the case $k=1$. Let $\{|e_i\ra:1\le i\le n\}$ be the canonical basis for $\IC^n$. Then $C(L)=\sum_{i,j=1}^n |e_i\ra\la e_j|\otimes L(|e_i\ra\la e_j|)$. We have
\begin{eqnarray*}
&&L\ge 0\\[2mm]
&\Lra &L(|y\ra \la y|)\ge 0\ \mbox{ for all }|y\ra\in \IC^n \\[2mm]
&\Lra& (\la y| \otimes I_{K}) C(L)(|y\ra\otimes I_{K})\ge 0 \ \mbox{ for all }|y\ra\in \IC^n \\[2mm]
&\Lra & \la z|\(\, (\la y\otimes I_{K}) C(L)(|y\ra\otimes I_{K})\, \) |z\ra\ge 0\ \mbox{ for all }|y\ra\in \IC^n,\ |z\ra\in K\\[2mm]
&\Lra & (\la y| \la z| ) C(L)(|y\ra|z\ra )\ge 0\ \mbox{ for all }|y\ra\in \IC^n,\ |z\ra\in K\\[2mm]
&\Lra&\la x| C(L)|x\ra \ge 0 \ \mbox{  for all }|x\ra =|y\ra |z\ra\mbox{ with }|y\ra\in \IC^n,\  |z\ra
\in  K.
\end{eqnarray*}

For general $k>1$, let $\{|f_p\ra:1\le p\le k\}$ be the canonical
basis for $\IC^k$. Then
\begin{equation}\label{CL}C(I_k\otimes L)=\sum_{p,q=1}^k\sum_{i,j=1}^n
\(|f_p\ra\la f_q|\otimes |e_i\ra\la e_j|)\otimes(|f_p\ra\la
f_q|\otimes L(|e_i\ra\la e_j|)\).\end{equation}
Note that every
$|\tilde y\ra \in \IC^{k}\otimes \IC^{n}$ (respectively, $|\tilde z\ra\in
\IC^{k}\otimes K$) has the form
\begin{equation}\label{yz}
|\tilde y\ra = \sum_{r=1}^k|f_r\ra\otimes |y_r\ra\quad \(\mbox{ respectively, } |\tilde z\ra=\sum_{s=1}^k|f_s\ra\otimes |z_s\ra\ \),\end{equation}
where $|y_r\ra \in \IC^n,\ |z_s\ra \in K,\ 1\le r,\, s\le k $.
\medskip
Now, applying the above result to $I_k\otimes L:M_{k}\otimes
M_{n}\to M_{k}\otimes B(K) $,   by  (\ref{CL}) and (\ref{yz}), we have
\begin{eqnarray*}
\hspace{-2cm}&& 
I_k\otimes L\ge 0\\[2mm]
\hspace{-2cm}&\Lra& 
(\la \tilde y|\otimes \la \tilde z|) C(I_k\otimes L)(|\tilde y\ra\otimes | \tilde z\ra) \ge 0
\mbox{  for all }|\tilde y\ra \in \IC^{k}\otimes \IC^{n}\mbox{ and } |\tilde z\ra\in \IC^{k}\otimes K\\[2mm]
\hspace{-2cm}&\Lra& 
\(\sum_{r,s=1}^k\la f_r|\la y_r| \la f_s | \la z_s |\)
\(\sum_{p,q=1}^k\sum_{i,j=1}^n
|f_p\ra\la f_q|\otimes |e_i\ra\la e_j|\otimes
|f_p\ra\la f_q|\otimes L(|e_i\ra\la e_j|\) \\[1mm]
\hspace{-2cm}&& 
\qquad \(\sum_{r',s'=1}^k |f_{r'}\ra |y_{r'}\ra |f_{s'}\ra  |z_{s'}\ra\) \ge 0
\mbox{  for all }|y_r\ra \in \IC^{n}\mbox{ and }|z_s\ra\in K, 1\le r,s\le k\\[1mm]
\hspace{-2cm}&\Lra& 
\(\sum_{p=1}^k \la y_p| \la z_p | \)\(\sum_{i,j=1}^n
|e_i\ra\la e_j| \otimes L(|e_i\ra\la e_j|)\)
\(\sum_{q=1}^k |y_{q}\ra |z_{q}\ra\) \ge 0 \\[1mm]
\hspace{-2cm}&& 
\qquad\mbox{  for all }| y_p\ra\in  \IC^{n}\mbox{ and }| z_p\ra\in K, 1\le p \le k \\[2mm]
\hspace{-2cm}&\Lra&
\la x|C(L)|x\ra \geq 0\mbox{ for all }|x\ra = \sum_{p=1}^k
|y_p\ra |z_p\ra\mbox{ with }|y_p\ra |z_p\ra \in \IC^n\otimes
K.
\end{eqnarray*}

\medskip
(b) $\Lra$ (c) : Suppose (c) holds. Given  $|x\ra=\sum_{p=1}^k|y_p\ra\otimes |z_p\ra$, where $|y_p\ra\in \IC^n$ and $|z_p\ra\in K$, $1\le p\le k$, let $P$ be the orthogonal projection to the subspace spanned by $\{|z_p\ra:1\le p\le k\}$. Then $(I_k\otimes P)|x\ra=|x\ra$. Therefore,
$$
\la x| C(L)|x\ra
= (\la x| (I_k\otimes P)) C(L)((I_k\otimes P) |x\ra ) =\la x|\(  (I_k\otimes P)C(L)(I_k\otimes P) \)|x\ra
\ge 0.$$
Conversely, suppose (b) holds. Let $P$ be an orthogonal projection in $K$ with rank $ k$
and $\{|z_p\ra:1\le p\le k\}$ be an orthonormal basis of the range space of $P$.
For every $|w\ra\in \IC^{n}\otimes K$, there exist $|y_p\ra\in \IC^n$, $1\le p\le k$ such that $(I_n\otimes P)|w\ra=\sum_{p=1}^k|y_p\ra\otimes |z_p\ra$. We have
$$\la w| (I_n\otimes P)C(L)(I_n\otimes P) |w\ra= \(\sum_{p=1}^k \la y_p| \la z_p| \) C(L) \(\sum_{p=1}^k |y_p\ra |z_p\ra \)\ge 0.$$
Hence, $(I_n\otimes P)C(L)(I_n\otimes P)\ge 0$.
\qed

\section{Elementary operators}

A linear map $L:B(H) \rightarrow B(K)$ is called an elementary
operator if it has the form
$$L(X) = \sum_{j=1}^k A_j X B_j^{\dag}$$
for some $A_1, \dots, A_k, B_1, \dots, B_k \in B(H,K)$ \cite{H2}. If
$H$ and $K$ are finite dimensional, then every linear map is
elementary. Since we are interested in positive linear map, we focus
on linear maps which map  self-adjoint operators to self-adjoint
operators. Thus, for any self-adjoint $X$,
$$\sum_{j=1}^k  A_jXB_j^{\dag} = L(X) = L(X)^{\dag} = \sum_{j=1}^k B_jXA_j^{\dag}.$$
As a result, for any self-adjoint $X$, we get
$$2L(X) =   \sum_{j=1}^k (A_jXB_j^{\dag} + B_jXA_j^{\dag})
= \sum_{j=1}^k (A_j+B_j)X(A_j+B_j)^{\dag} -\sum_{j=1}^k
(A_jXA_j^{\dag} + B_jXB_j^{\dag}).$$ By linearity, the above
equation is true for all $X \in B(H)$. Thus we will focus on
elementary operators of the form
$$L(X) = \sum_{j=1}^p C_jXC_j^{\dag} - \sum_{j=1}^q D_j X D_j^{\dag}.$$
Hou \cite{H2} gave a condition for an elementary operator in the
above form to be $k$-positive. In this section, we will extend those
results by Proposition 2.1 in the following.

\begin{proposition} \label{3.1}
Suppose $L:B(H) \to B(K)$ has the form
\begin{equation}\label{elementary}
X \mapsto \sum_{r=1}^p C_rXC_r^{\dag} - \sum_{s=1}^q D_s X D_s^{\dag}
\end{equation}
with $C_1, \dots, C_p, D_1, \dots, D_q\in B(H,K)$.
Then $$(I_k\otimes L)(X) = \sum_{r=1}^p (I_k\otimes C_r) X (I_k \otimes C_r^{\dag}) -
\sum_{s=1}^q (I_k \otimes D_s) X (I_k \otimes D_s^{\dag}).$$
Moreover, the following are equivalent.
\begin{enumerate}
\item[{\rm (a)}] $L$ is $k$-positive, i.e., $I_k \otimes L$ is positive.

\item[{\rm (b)}] $\sum_{r=1}^p (I_k\otimes C_r) X (I_k \otimes C_r^{\dag}) -
\sum_{s=1}^q (I_k \otimes D_s) X (I_k \otimes D_s^{\dag}) \in M_k(B(K))^+$ for
any rank one orthogonal projection $X \in M_k(B(H))$.

\item[{\rm (c)}] For any (orthonormal) subset $\{|x_1\ra, \dots, |x_k\ra\}
\subseteq H$, $\sum_{i,j=1}^k E_{ij} \otimes L(|x_i\ra \la x_j|)$ is positive
semi-definite, equivalently,
$$\sum_{r=1}^p \sum_{i,j=1}^k E_{ij} \otimes C_r |x_i\ra \la
x_j|C_r^{\dag} \ge \sum_{s=1}^q \sum_{i,j=1}^k E_{ij} \otimes D_s |x_i\ra \la
x_j|D_s^{\dag}.$$

\item[{\rm (d)}] For any $| x \ra\in \IC ^k\otimes H$, there is an $q\times
p$ matrix $T_x$ with operator norm $\|T_x\|\leq 1$ such that
$$\left(\begin{array}{c} I_k\otimes D_1\\I_k\otimes D_2\\ \vdots \\
I_k\otimes D_q\end{array}\right)|x\ra
=(T_x\otimes I_K)\left(\begin{array}{c} I_k\otimes C_1\\I_k\otimes C_2\\
\vdots \\ I_k\otimes C_p\end{array}\right)|x\ra.$$
\end{enumerate}

\end{proposition}

\it Proof. \rm The equivalence of  (b) and (c) follows from
Proposition \ref{2.1} and the special form of $L$.  For the
equivalence of (a), (b) and (d), see \cite{H2}. \qed

Recall that, for a linear operator $A\in B(H)$ and a positive
integer $k \le \dim H$,  the $k$-numerical range of $A$ is defined
by
$$W_k(A) = \left\{\sum_{j=1}^k \la x_j|A|x_j\ra: \{|x_1\ra, \dots, |x_k\ra\}
\mbox{ is an orthonormal set in } H\right\}.$$ If $\dim H=n <
\infty$, and $A$ is Hermitian with eigenvalues $a_1 \ge \cdots \ge
a_n$, then
$$W_k(A) = \left[\sum_{j=1}^k a_{n-j+1}, \sum_{j=1}^k a_j\right].$$
For the details of $k$-numerical ranges, see \cite{B}.

The following proposition gives the relation between $k$-numerical
ranges and $k$-positivity of elementary operators.

\begin{proposition}
\label{3.2} Suppose $L:M_n\rightarrow B(K)$ has the form (\ref{elementary}).
\begin{enumerate}

\item[{\rm (a)}] If $L$ is $k$-positive, then $W_k \left(\sum_{r=1}^p
C_r^{\dag}C_r-\sum_{s=1}^q D_s^{\dag} D_s \right) \subseteq [0, \infty)$.

\item[{\rm (b)}] If for any unit vectors $|u\ra = (u_1, \dots, u_p)^t\in \IC^p$
and $|v\ra = (v_1, \dots, v_q)^t \in \IC^q$,
\begin{equation}\label{wk-inequality}
\hspace{-2cm}\min W_k\left((\sum_r u_rC_r)^{\dag}(\sum_r u_rC_r)\right)\ge
\max W_k\left((\sum_s v_sD_s)^{\dag}(\sum_s v_sD_s)\right),
\end{equation}
then  $L$ is $k$-positive.
\end{enumerate}
Here, $\min S$ and $\max S$ denote the minimum and maximum value of a subset $S$ of real number.
\end{proposition}

\it Proof. \rm Denote by $\Gamma_{n,k}$ the set of vectors $|\bx\ra
= \pmatrix{ |x_1\ra \cr \vdots \cr |x_k\ra\cr}$ such that
$\{|x_1\ra, \dots, |x_k\ra\}\subseteq \IC^n$ is an orthonormal set.

If $L$ is $k$-positive, then $(I_k\otimes L)(|\bx\ra \la \bx|)$ is positive semi-definite
for every  $|\bx\ra \in \Gamma_{n,k}$. Taking trace, we see that
$$0 \le \sum_{j=1}^k{\rm tr}\left(\sum_r C_r |x_j\ra \la x_j|C_r^{\dag} -
\sum_s D_s |x_j\ra \la x_j| D_s^{\dag}\right) = \sum_{j=1}^k \la
x_j| \left(\sum_r C_r^{\dag}C_r - \sum_s D_s^{\dag}D_s\right)
|x_j\ra.$$ The result (a) follows.

For (b), suppose (\ref{wk-inequality}) holds for any unit vectors
$|u\ra = (u_1, \dots, u_p)^t \in \IC^p$ and $|v\ra = (v_1, \dots,
v_q)^t\in \IC^q$. For $|\bx\ra \in \Gamma_{n,k}$, let
$$\tilde C_{\bx} =
\pmatrix{C_1|x_1\ra & \cdots & C_p|x_1\ra \cr \vdots & \ddots & \vdots \cr
C_1|x_k\ra & \cdots &C_p |x_k\ra\cr}
\quad \hbox{ and } \quad
\tilde D_{\bx} =
\pmatrix{D_1|x_1\ra & \cdots & D_q|x_1\ra \cr \vdots & \ddots & \vdots \cr
D_1|x_k\ra & \cdots &D_q |x_k\ra\cr}.$$
We will show that $\tilde C_{\bx}\tilde C_{\bx}^{\dag}
- \tilde D_{\bx} \tilde D_{\bx}^{\dag}$
is positive semi-definite, or equivalently, for any unit vector $|y\ra \in \IC^{kp}$,
$$\|\la y| \tilde C_{\bx}\|^2 \ge \|\la y| \tilde D_{\bx}\|^2.$$

Denote by $\sigma_1(A) \ge \sigma_2(A) \ge \cdots \ge \sigma_{\min\{m,n\}}(A)$
be the singular values of $A \in M_{m,n}$.
Note that there is $|\tilde \bx\ra\in \Gamma_{n,k}$ so that $\tilde
C_{\tilde\bx}$ has the smallest $p$-th singular value $\sigma_p(\tilde C_{\tilde\bx})$ among
all choices of $|\bx\ra \in \Gamma_{n,k}$.
$$\|\la y| \tilde C_{\bx} \| \ge \sigma_{p}(\tilde C_{\bx}) \ge \sigma_p(\tilde
C_{\tilde\bx}).$$
Moreover, there is a unit vector $|\tilde  u\ra  = (\tilde u_1, \dots, \tilde u_p)^t \in \IC^p$
such that
\begin{eqnarray*}
(\sigma(\tilde
C_{\tilde\bx}))^2
= \|\tilde C_{\tilde \bx} |\tilde  u\ra\|^2
&=& \left\| \pmatrix{(\sum_r \tilde u_r C_r) |\tilde x_1\ra \cr \vdots \cr (\sum_r \tilde u_r C_r)
|\tilde x_k\ra \cr}\right\|^2 \\
&=& \sum_{j=1}^k \la \tilde x_j| (\sum_r \tilde u_r C_r)^{\dag}(\sum_r \tilde u_r C_r) |\tilde x_j\ra \\
&\ge&  \min W_k\left((\sum_r \tilde u_r C_r)^{\dag}(\sum_r \tilde  u_r C_r)\right).
\end{eqnarray*}
Similarly, we can choose $|\hat \bx\ra\in \Gamma_{n,k}$ so that
$\tilde D_{\hat \bx}$ has the largest maximum singular value $\sigma_1(\tilde D_{\hat \bx})$
among all choice of $|\bx\ra\in \Gamma_{n,k}$.
Then
$$\|\la\by| \tilde D_{\bx}\| \le \sigma_{1}(\tilde D_{\bx}) \le \hat \sigma_1(\tilde D_{\hat \bx}).$$
Moreover, there is a unit vector $|\hat   v\ra = (\hat v_1, \dots, \hat v_q)^t \in \IC^q$
such that
\begin{eqnarray*}
\hspace{-2.2cm} 
\max W_k\left((\sum_s v_s D_s)^{\dag}(\sum_s v_s D_s)\right) &\ge& 
\sum_{j=1}^k \la \hat x_j| (\sum_s \hat v_s D_s)^{\dag}(\sum_s \hat v_s D_s) |\hat x_j\ra \\
\hspace{-2cm} &=&  \left\| \pmatrix{(\sum_s \hat v_s D_s) |\hat x_1\ra \cr \vdots \cr (\sum_s \hat v_s D_s)
|\hat x_k\ra\cr}\right\|^2  =  \|\tilde D_{\hat \bx} |\hat  v\ra\|^2 =  (\sigma_1(\tilde D_{\hat \bx}))^2.
\end{eqnarray*}
By our assumption, we have $\sigma_p(\tilde C_{\tilde \bx}) \ge  \sigma_1(\tilde D_{\hat \bx})$, and hence
$$\|\la  y| \tilde C_{\bx}\| \ge \sigma_p(\tilde C_{\tilde \bx}) \ge  \sigma_1(\tilde D_{\hat \bx}) \ge \|\la y| \tilde D_{\bx}\|.$$
The desired conclusion follows.
\qed

\medskip
\noindent {\bf Remark} Note that in the above proof, if there is
$|\bx\ra \in \Gamma_{n,k}$ such that $\tilde C_{\bx} = 0$, then
$$\min\left\{W_k\left((\sum_r u_rC_r)^{\dag}(\sum_r u_rC_r)\right): | u\ra
=(u_1, \dots, u_p)^t \in \IC^p, \ \la u| u\ra = 1\right\} = 0.$$
On the other hand, if
$$\min\left\{W_k\left((\sum_r u_rC_r)^{\dag}(\sum_r u_rC_r)\right): | u\ra=
(u_1, \dots, u_p)^t \in \IC^p, \ \la  u| u\ra = 1\right\} > 0,$$
then $\tilde C_{\bx}$ has rank $kp$ for all
$|\bx\ra \in \Gamma_{n,k}$.

\section
{Clarification and improvement of some results of Chru\'{s}ci\'{n}ski and  Kossakowski}

In this section, we give a numerical range criterion for $k$-positivity of 
maps $\phi:M_{n}\rightarrow M_{m}$ defined by
$$L(X) = L_1(X)- L_2(X)$$
with
$$L_1(X) = \sum_{j=1}^p \gamma_j F_jXF_j^{\dag} \quad \hbox{and}\quad
L_2(X) =  \sum_{j=p+1}^{mn} \gamma_j F_jXF_j^{\dag}, \qquad
\gamma_1, \dots, \gamma_{mn} \ge 0,$$
where $F_1, \dots, F_{mn} \in M_{m,n}$ satisfy the following:

\medskip\noindent
{\bf (C1)} \it $\{F_j\}_{j=1}^{mn} \subseteq
M_{m,n}$ is an orthonormal set using the inner product $(X,Y) =
\tr(XY^{\dag})$. \rm

\medskip
The positivity of such maps were also considered in \cite{CK,CK2} using the norm
$$\|X\|_k = \left\{\sum_{j=1}^k \sigma_j(X)^2 \right\}^{1/2},$$
where $\sigma_1(X) \ge \sigma_2(X) \ge \cdots \ge \sigma_{\min\{m,n\}}(X)$ are 
the singular values of $X \in M_{m,n}$.
The norm $\|X\|_k$ is known as the $(2,k)$-spectral norm; see \cite{Li}.
The authors of \cite{CK} mistakenly referred this as the Ky Fan $k$-norm 
$|X|_k = \sum_{j=1}^k \sigma_j(X)$ of the matrix $X$; see \cite[p.445]{Horn}.

\medskip
Consider the following condition on $\{F_j\}_{j=1}^{mn} \subseteq M_{mn}$.

\medskip\noindent
{\bf (C2)} \it For any orthonormal basis
$\{|x_1\rangle,|x_2\rangle,\ldots, |x_n\rangle\}$ of $\IC^n$,
$$\left\{P_k= \sum_{i,j=1}^n |x_i\ra\la x_j|\otimes
F_k|x_i\ra \la x_j|F_k^\dag: 1 \le k \le mn\right\}$$ 

is  a set of mutually orthogonal set of rank one matrices.

\medskip\noindent
\rm
In  \cite{CK}, the authors showed that under the assumption {\bf (C2)},
if $1 > \sum_{j=p+1}^{mn}\|F_j\|_k^2$ and
$$\sum_{i,j=1}^n E_{ij}\otimes L_1(E_{ij})
\geq \frac{\sum_{j=p+1}^{mn}\gamma_j\|F_j\|_k^2}
{1-\sum_{j=p+1}^{mn}\|F_j\|_k^2}
\left( I_n\otimes I_m-\sum_{j=p+1}^{mn}P_j \right),$$
then $\phi$ is $k$-positive. 

In \cite{CK2}, the authors stated the result using the assumption 
{\bf (C1)} instead of {\bf (C2)} without explaining their relations.
In addition, there are some typos in the papers that further obscured
the results.

In the following, we use 
results in the previous sections to refine and improve the
results in \cite{CK,CK2}. We first show that condition {\bf (C1)} and 
{\bf (C2)} are equivalent.  Note that 
the special case for $m = n$ was treated in \cite[Lemma 1]{SS}.

\begin{proposition} \label{4.1}
Suppose $\{F_1, \dots, F_{mn}\} \subseteq M_{m,n}$. The following two conditions are equivalent.
\begin{enumerate}
\item[{\rm (a)}] $\{F_1, \dots, F_{mn}\}$ is an orthonormal set, i.e. $\tr(F_r^\dag F_s) =\delta_{r\,s}$ for $r, s=1,\dots mn$.

\item[{\rm (b)}]  For any orthonormal basis $\{|x_1\ra, \dots, |x_n\ra\}$
of $\IC^n$,
$$\left\{ \sum_{i,j=1}^n |x_i\ra \la x_j| \otimes F_r|x_i\ra \la x_j|F_r^{\dag}  : 1 \le r \le mn \right\}$$
is a set of mutually orthogonal rank one
projections in $M_{mn}$.

\end{enumerate}
\noindent
Furthermore, if (a) or (b) holds, then we have
$$\sum_{j} F_jF_j^{\dag} = nI_m\mbox{ and }\sum_{j} F_j^{\dag}F_j = mI_n\,.$$

\end{proposition}

\it Proof. \rm 
Suppose $ F_1, \dots, F_{mn}\in M_{m,n}$ and $\{|x_1\ra, \dots, |x_n\ra\}$ is an orthonormal basis in $\IC^n$. Define 
$$P_r=  \sum_{i,j=1}^n |x_i\ra \la x_j| \otimes F_r|x_i\ra \la x_j|F_r^{\dag} \quad\hbox{for } r=1,\dots, mn.$$
Then
for any $r,s=1,\dots, mn$, we have
\begin{eqnarray*}
P_r P_s 
&=& 
\(\sum_{i,j=1}^n |x_i\ra \la x_j| \otimes F_r|x_i\ra \la x_j|F_r^{\dag} \)\(\sum_{k,\ell=1}^n |x_k\ra \la x_\ell| \otimes F_s|x_k\ra \la x_\ell|F_s^{\dag} \) \\
&=& \sum_{i,j,k,\ell = 1}^n |x_i\ra \la x_j| x_k \ra \la x_\ell|  \otimes 
F_r|x_i\ra \la x_j|F_r^{\dag}
F_s|x_k\ra \la x_\ell |F_s^{\dag} \\
&=& \left(\sum_{j,k=1}^n \la x_j | x_k \ra \cdot \la x_j|F_r^{\dag}
F_s|x_k\ra \right)\left( \sum_{i,\ell = 1}^n    |x_i\ra \la x_\ell|  \otimes 
F_r|x_i\ra \la x_\ell |F_s^{\dag}  \right) \\
&=& \left(\sum_{j=1}^n \la x_j|F_r^{\dag}
F_s|x_j\ra \right)\left( \sum_{i,\ell = 1}^n    |x_i\ra \la x_\ell|  \otimes 
F_r|x_i\ra \la x_\ell |F_s^{\dag}  \right) \\
&=& \tr( F_r^\dag F_s ) 
\left( \sum_{i,\ell = 1}^n    |x_i\ra \la x_\ell|  \otimes 
F_r|x_i\ra \la x_\ell |F_s^{\dag}  \right).
\end{eqnarray*}
Therefore, the implication (a) $\Rightarrow$ (b) holds. Now by taking the trace on both sides of the equation,
\begin{eqnarray*}
\tr(P_rP_s)
&=& \tr( F_r^\dag F_s ) \cdot  
\tr\left( \sum_{i,\ell = 1}^n    |x_i\ra \la x_\ell|  \otimes 
F_r|x_i\ra \la x_\ell |F_s^{\dag}  \right) \\
&=& \tr( F_r^\dag F_s ) \cdot  
\left( \sum_{i,\ell = 1}^n    \la x_\ell|x_i\ra \cdot \la x_\ell| F_s^\dag F_r|x_i\ra   \right) \\
&=& \tr( F_r^\dag F_s ) \cdot  
\left( \sum_{i= 1}^n    \la x_i| F_s^\dag F_r|x_i\ra   \right) \\
&=& \tr( F_r^\dag F_s ) \cdot \tr ( F_s^\dag F_r) = | \tr (F_r^\dag F_s) |^2.
\end{eqnarray*}
Hence, (b) $\Rightarrow$ (a) followed by the above equlity. 

Finally suppose (a) holds. We have
$$\tr (F_iF_j^{\dag})=\tr (F_j^{\dag}F_i) =\delta_{ij},\ \mbox{for all }1\le j\le mn.$$

 For $1\le j \le mn$ and $1\le r\le m$, let $f_j^r$ be the
$r^{\rm th}$ row of $F_j$. From $\tr F_iF_j^{\dag}=\delta_{ij}$, we
can form a unitary matrix $U  \in M_{mn}$ with $j^{\rm th} $ row
$u_j=\left[\, f_j^1\, |\, \cdots\, |\, f_j^m\, \right]$.  Since $U^{\dag}U=I_{mn}$,  for $1\le
r,\ s\le m$, we have
$\sum_{j=1}^{mn}(f_j^s)^{\dag}f_j^{r}=\delta_{rs}I_n$. Consider  $R
= \sum_{j=1}^{mn} F_jF_j^{\dag} \in M_m$. The $(r,s)$-th entry of $R$
is equal to
$$\sum_{j=1}^{mn}f_j^{r}(f_j^s)^{\dag}=\tr \(\sum_{j=1}^{mn}f_j^{r}(f_j^s)^{\dag}\)
=\tr \(\sum_{j=1}^{mn}(f_j^s)^{\dag}f_j^{r}\)=\tr (\delta_{rs}I_n)=n\delta_{rs}\,.$$
Therefore,  $\sum_{j} F_jF_j^{\dag} = nI_m$. Similarly, by replacing $F_j$
with $F_j^{\dag}$, $\sum_{j} F_j^{\dag}F_j = mI_n$ follows from
the fact that $\tr F_j^{\dag} F_i=\delta_{ij}$ for all $1\le i,j\le mn$.
\qed

Using the concept of the $k$-numerical range, we have the following.

\begin{proposition} \label{4.3} Suppose that $\{F_{j}:1 \le j \le mn\}$ is an orthonormal basis
of $M_{m,n}$ and  $L:M_{n} \rightarrow M_m$ has the form
$$L(X) = \sum_{j=1}^p \gamma_j F_jXF_j^{\dag} - \sum_{j=p+1}^{mn} \gamma_j F_jXF_j^{\dag}, \quad
\gamma_1, \dots, \gamma_{mn} \ge 0.$$
Assume that $1 \le k \le \min\{m,n\}$ and
$\xi_k = 1 - \max  W_k (\sum_{j=p+1}^{mn} F_j^{\dag}F_j) > 0$.
\begin{enumerate}
\item[{\rm (a)}]
If
$$\gamma_i \ge
\xi_k^{-1}\, \max W_k\left(\sum_{j=p+1}^{mn}\gamma_j F_j^{\dag}F_j\right),
\qquad i = 1, \dots, p,$$ then $L$ is $k$-positive.

\item[{\rm (b)}] If $p = mn-1$ and
$$\gamma_i <
\xi_k^{-1} \gamma_{mn} \max W_k\left(F_{mn}^{\dag}F_{mn}\right) = \gamma_{mn}\|F_{mn}\|_k^2 
\qquad \hbox{for all} \quad i = 1, \dots, mn-1,$$
then $L$ is not $k$-positive.
\end{enumerate}
\end{proposition}

\it Proof. \rm 
(a) Let $w_k = \max W_k\left(\sum_{r=p+1}^{mn} \gamma_r F_r^{\dag}F_r\right)$. 
Suppose $ \gamma_i \ge \xi_k^{-1} w_k$ for each $i = 1, \dots, p$.
Denote by $\Gamma_{n,k}$ the set of
vectors $|\bx\ra$ such that $|\bx\ra = \pmatrix{|x_1\ra\cr \vdots \cr |x_k\ra}$
where $\{|x_1\ra , \dots, |x_k\ra\}$ is an orthonormal set in $\IC^n$.
We show that $I_k\otimes L(|\bx\ra \la \bx|)$ is positive
semidefinite for any $|\bx\ra \in \Gamma_{n,k}$. The conclusion will then 
follow from Proposition \ref{3.1}.

We may extend $|\bx\ra \in \Gamma_{n,k}$ 
to $|\tilde \bx\ra \in \Gamma_{n,n}$  with $|\tilde \bx\ra
= \pmatrix{|x_1\ra \cr \vdots \cr |x_n \ra}$ such
that $\{|x_1\ra, \dots, |x_n\ra\}$ is an orthonormal basis for $\IC^n$. By
Proposition \ref{4.1}(b),
$$\sum_{r=1}^{mn}(F_{r}|x_i\ra \la x_j|F_{r}^{\dag})_{1 \le i, j \le n} = I_{mn}.$$
Focusing on the leading $mk \times mk$ principal submatrix, we have
\begin{equation}
\label{I-mk}
\sum_{r=1}^p (F_{r}|x_i\ra\la x_j|F_{r}^{\dag})_{1 \le i, j \le k}
= I_{mk} - \sum_{r=p+1}^{mn}(F_{r}|x_i\ra \la x_j|F_{r}^{\dag})_{1 \le i, j \le k}.
\end{equation}
Note that
$$\tr\left(\sum_{r=p+1}^{mn} \gamma_r(F_r |x_i\ra\la x_j|F_r^{\dag})_{1 \le i, j \le k}\right)
= \sum_{j=1}^k \la x_j|\left(\sum_{r=p+1}^{mn}\gamma_r F_r^{\dag}F_r\right)|x_j\ra \le w_k.$$
Thus,
\begin{equation}\label{wk-1}
\sum_{r=p+1}^{mn} \gamma_r(F_r |x_i\ra\la x_j|F_r^{\dag})_{1 \le i, j \le k} 
\le w_k I_{mk}.
\end{equation}
Applying this argument to the special case when $\gamma_{p+1} = \cdots = \gamma_{mn}$,
we see that
$$\sum_{r=p+1}^{mn} (F_r |x_i\ra\la x_j|F_r^{\dag})_{1 \le i, j \le k}
\le \max W_k\left(\sum_{j=p+1}^{mn} F_j^*F_j\right) I_{mk} .$$
By (\ref{I-mk}) and the fact that $\xi_k 
= 1-\max W_k\left(\sum_{j=p+1}^{mn} F_j^*F_j\right)$, we have
$$\xi_k I_{mk} \le \sum_{r=1}^{p} (F_r |x_i\ra\la x_j|F_r^{\dag})_{1 \le i, j \le k}.$$
Because
$\gamma \xi_k^{-1} \leq \gamma_i$ for each $i = 1, \dots, p$, we have 
\begin{equation}\label{wk-2}
\hspace*{-1cm}
w_k I_{mk} \le  w_k \xi_k^{-1}
\sum_{r=1}^{p} (F_r |x_i\ra\la x_j|F_r^{\dag})_{1 \le i, j \le k} \le
\sum_{r=1}^{p} \gamma_r(F_r|x_i\ra \la x_j|F_r^{\dag})_{1 \le i, j \le k}.
\end{equation}
By (\ref{wk-1}) and (\ref{wk-2}), we have the desired operator inequality
$$
\sum_{r=p+1}^{mn} 
\gamma_r(F_r |x_i\ra\la x_j|F_r^{\dag})_{1 \le i, j \le k} 
\le \sum_{r=1}^{p} \gamma_r(F_r|x_i\ra \la x_j|F_r^{\dag})_{1 \le i, j \le k}.
$$

\medskip
(b) Suppose that the hypothesis of (b) holds. We can choose
$|x_1\ra, \dots, |x_k\ra$ in $\IC^n$ so that
$$\tr \, (F_{mn}|x_i\ra \la x_j|F_{mn}^{\dag})_{1\le i,j\le k}  = 
\max W\left(F_{mn}^{\dag}F_{mn}\right) = \|F_{mn}\|_k^2,$$
i.e., the rank one matrix $\gamma_{mn}(F_{mn}|x_i\ra\la x_j|F_{mn}^{\dag})_{1\le i,j\le k}$ has a nonzero eigenvalue
$\gamma_{mn} \|F_{mn}\|_k^2$.
Now,
$$ \sum_{r=1}^{mn-1} \gamma_r(F_r |x_i\ra\la x_j|F_r^{\dag})_{1\le i,j\le k}
<   \xi_k^{-1} \gamma_{mn}  \left(\sum_{r=1}^{mn-1}(F_r |x_i\ra\la x_j|F_r^{\dag})_{1\le i,j\le k}\right)
\le  \gamma_{mn}I_{mk}.$$
Thus, the matrix
$$\sum_{r=1}^{mn-1} \gamma_r(F_r |x_i\ra\la x_j|F_r^{\dag})_{1\le i,j\le k}
- \gamma_{mn}(F_{mn} |x_i\ra\la x_j|F_{mn}^{\dag})_{1\le i,j\le k}$$
has a negative eigenvalue. The result follows from Proposition \ref{3.1}.
\qed

Part (b) of the Proposition \ref{4.3} was proved in \cite{CK,CK2}.
Using the fact that 
$$\max W_k\left(\sum_{r=p+1}^{mn} \gamma_r F_r^{\dag}F_r\right)
\le \sum_{r= p+1}^{mn} \gamma_r \|F_r\|_k^2$$
for any nonnegative numbers $\gamma_{p+1}, \dots, \gamma_{mn}$, we can
deduce the main result in \cite{CK,CK2}.

\begin{corollary} \label{4.2} Suppose $\{F_{j}:1 \le j \le mn\}$ is an orthonormal
basis of $M_{m,n}$ and  $L:M_{n} \rightarrow M_m$ has the form
$$L(X) = \sum_{j=1}^p \gamma_j F_jXF_j^{\dag} - \sum_{j=p+1}^{mn} \gamma_j F_jXF_j^{\dag},
\quad \gamma_1, \dots, \gamma_{mn} \ge 0.$$
Assume that $1 \le k \le \min\{m,n\}$ and  $\tilde \xi_k = 1 - \sum_{j=p+1}^{mn} \|F_j\|_k^2 > 0$.
If
$$\gamma_i \ge
\tilde \xi_k^{-1} \left(\sum_{j=p+1}^{mn} \gamma_j\|F_j\|_k^2\right) 
\qquad \hbox{for all}\quad i = 1, \dots, p,$$
then $L$ is $k$-positive.
\end{corollary}

\medskip
In \cite{CK2}, the authors mistakenly claimed that if $\sum_{j = p+1}^{mn} \|F_j\|_{k+1}^2 < 1$ and
$$\gamma_i < \tilde \xi_{k+1}^{-1} \left( \sum_{j=p+1}^{mn} \gamma_j \|F_j\|_{k+1}^2\right) \quad \hbox{for all } i =1,\dots,p,$$
then $L$ is not $(k+1)$-positive, see \cite[Theorem 1]{CK2}. 
The case when $p = mn=1$ was proved in \cite[Theorem 1]{CK}
and also Proposition \ref{4.3}. 
However, the following example shows that 
the claim in general does not hold when $p < mn-1$.
Also, this example demonstrates that Proposition \ref{4.3} is stronger than Corollary \ref{4.2}
(the result in \cite{CK,CK2}).

\begin{example} \rm
Let $m=n=8$.
Suppose $\{F_1, \dots, F_{64}\}$ is an orthonormal basis for $M_{8}$ 
such that 
$$F_{63} = \frac{1}{4} I_4 \otimes \pmatrix{1 & 1 \cr 1 & 1 \cr} 
\quad \hbox{ and } \quad 
F_{64} = \frac{1}{4} I_4 \otimes \pmatrix{1 & -1 \cr -1 & 1 \cr}.$$ 
Define $L: M_8 \rightarrow M_8$ by
$$L(X) =  \sum_{j=1}^{62} \gamma_j F_jXF_j^{\dag} - 
\sum_{j=63}^{64} F_jXF_j^{\dag}.$$ 
Then $\|F_{63}\|_2^2 + |F_{64}\|_2^2 = 1/2 + 1/2 = 1$ and 
$W_2\left(\sum_{j=63}^{64} F_j^{\dag}F_j\right) = W_2(I_8/4) = \{1/2\}$.
So, Corollary \ref{4.2} is not applicable.
By Proposition \ref{4.3}, the map $L$ is 2-positive if
$\gamma_i \ge 1$ for $i = 1, \dots, 62$.
\end{example}

\section{Criteria for $k$-positivity of $D$-type linear maps}

In this section, we consider linear maps  $L:M_n \rightarrow M_n$ of
the form
\begin{equation}\label{map}
A = (a_{ij})\mapsto \diag \left(\sum_{k=1}^n a_{kk} d_{k1},\, \dots,\, \sum_{k=1}^n a_{kk} d_{kn} \right) -  A
\end{equation}
for an $n\times n$ nonnegative matrix $D = (d_{ij})$. This type of
maps will be called $D$-type linear maps. The question of when a
$D$-type map is positive was studied intensively by many authors and
applied in quantum information theory to detect entangled states and
construct entanglement witnesses.  For example, if $D =
(n-1)I_n + E_{12} + \cdots + E_{n-1,n} + E_{n,1}$, we get a
positive map which is not completely positive.
This can be viewed as a generalization of the Choi map in \cite{Choi}.

In the following, we present a necessary and sufficient
criteria of $D$-type linear map to be $k$-positive.

\begin{proposition} \label{5.1} Suppose $L: M_n \rightarrow M_n$
is a $D$-type map  (\ref{map}) for an $n\times n$ nonnegative matrix
$D = (d_{ij})$. The following conditions are equivalent.
\begin{itemize}
\item[{\rm (a)}] $L$ is $k$-positive.
\item[{\rm (b)}] $d_{jj}>0$ for all $j=1,\dots,n$, 
and for any $k\times n$ matrix 
$U$ with columns $|u_1\ra, \dots, |u_n\ra \in \IC^k$
satisfying $\tr(U^{\dag}U) = 1$, we have
$$\sum_{j=1}^n \la u_j|\, (U \diag(d_{1j}, \dots, d_{nj}) U^{\dag})^{[-1]}\, 
|u_j\ra \le 1,$$
where $X^{[-1]}$ is the Moore-Penrose generalized
inverse of $X$.
\end{itemize}
\end{proposition}

\it Proof. \rm
For any $k\times n$ matrix $U$ with column $|u_1\ra, \dots, |u_n\ra \in \IC^k$ 
satisfying $\tr(U^{\dag}U) = 1$, consider the unit vector
$$|\bu \ra = \sum_{j=1}^n |u_j \ra \otimes |\hat e_j\ra \in \IC^{nk},$$
which can also be expressed as
$$|\bu \ra  = \sum_{j=1}^k |e_j \ra \otimes |\hat u_j\ra,$$
where $\{|e_1\ra,\dots,|e_k\ra\}$ and $\{|\hat e_1\ra,\dots,|\hat e_n\ra \}$ 
are the standard basis of $\IC^k$ and $\IC^n$, respectively, and
$|\hat u_j\ra$ is the transpose of the $j$-th row of $U$. Let
$$F_{ij} = L(|\hat u_i\ra \la \hat u_j|)+|\hat u_i\ra\la \hat u_j|,$$
which is a diagonal matrix with the $\ell $-th diagonal entry equal to 
$\la \hat u_j |\diag(d_{1\ell},\dots,d_{n\ell}) |\hat u_i\ra$.
By Proposition \ref{2.1}, $L$ is $k$-positive if and only if the $nk \times nk$ matrix
$$(I_k\otimes L)(|\bu\ra\la\bu|) = 
\sum_{i,j=1}^k |e_i\ra \la e_j| \otimes \(F_{ij} -| \hat u_i \ra\la \hat u_j|\) 
= \(\sum_{i,j=1}^k |e_i\ra \la e_j| \otimes F_{ij}\) -|\bu\ra \la \bu|$$
is positive semi-definite. Notice that the above matrix is permutationally similar to
$$\sum_{i,j=1}^k  \(F_{ij} -| \hat u_i \ra\la \hat u_j|\) \otimes |e_i\ra\la e_j|
= \(\sum_{i,j=1}^k F_{ij} \otimes |e_i\ra\la e_j|\) - |\hat \bu \ra\la \hat \bu|,$$
where $|\hat \bu \ra = \sum_{j=1}^n |\hat e_j\ra \otimes |u_j\ra$. 
Direct computation shows that
$$\sum_{i,j=1}^k F_{ij} \otimes E_{ij} = D_1 \oplus \cdots \oplus D_n,$$
where
$D_j = U\, \diag(d_{1j}, \dots, d_{nj}) U^{\dag}$
for $j = 1, \dots, n$.
Therefore, the condition is equivalent to:

\medskip
(c) $|\hat \bu\ra$
lies in the range of $D_1\oplus \cdots \oplus D_n$ and $\| (D_1\oplus \cdots \oplus D_n)^{1/2})^{[-1]} |\hat \bu\ra\| \le 1$.

\medskip\noindent
Note that for any choice of $U$ satisfying $\tr(U^{\dag}U) = 1$
the corresponding $|\hat u_j\ra $ always lies in the range
of $D_j$ for $j = 1, \dots, n$ if and only if  
$d_{jj} > 0$ for all $j = 1, \dots, n$;
the norm inequality in (c) is the same as the inequality stated in (b).
Therefore, condition (a) is equivalent to condition (c), which is equivalent to 
condition (b).
\qed

\medskip\noindent

Proposition \ref{5.1} is particularly useful when $k=1$.

\begin{corollary} \label{5.2} Let $L: M_n \rightarrow M_m$ be a $D$-type map of the
form  (\ref{map}) with $D=(d_{ij})$. For $u=(u_1, u_2,\ldots, u_n)^t \in \IC^{n}$,
let $f_j(u)=\sum_{i=1}^n d_{ij}|u_i|^2 $.
Then, $L$ is positive if and only if any one of the following equivalent conditions hold
\begin{itemize}
\item[$(1) $]  $d_{ii}>0$ for all $i=1,\dots,n$ and  $ \sum_{u_j \ne 0} \frac{|u_j|^2}{f_{j}(u)}\leq 1$
 for every unit
vector $|\bu\ra=(u_1, u_2,\ldots, u_n)^t \in \IC^{n}$.
\item[$(2) $]  $d_{ii}>0$ for all $i=1,\dots,n$ and  $ \sum_{j =1}^n \frac{|u_j|^2}{f_{j}(u)}\leq 1$
 for every
vector $|\bu\ra=(u_1, u_2,\ldots, u_n)^t \in \IC^{n}$ with $u_i\ne 0$ for all  $i=1,\dots,n$.

\end{itemize}
\end{corollary}
\it Proof. \rm For $k=1$, (1) is equivalent to condition (b) in Proposition \ref{5.1}. (2) is equivalent to (1) because $\frac{|u_j|^2}{f_{j}(u)}$ is   continuous and homogeneous in $u$.
\qed

\section{Constructing $D$-type positive maps}

In this section, we discuss how to construct $D$-type positive
linear maps using the results in previous sections.

The following example is well-known. Here we give a different proof
by applying Proposition \ref{2.1}.
Notice that the map is a $D$-type map with all entries of $D$ being $\gamma$.

\begin{example} \label{6.1}
For $\gamma \ge 0$, define $L_{\gamma}: M_n\rightarrow M_n$ by
$$L_\gamma(A) = \gamma(\tr A)I_n - A.$$
Then for any $k \in \{1, \dots, n\}$, $L_\gamma$ is
$k$-positive if and only if $\gamma \ge k$.
\end{example}

\it Proof. \rm For any $|x\ra = \sum_{j=1}^k |e_j\ra \otimes |x_j\ra
\in \IC^{nk}$ with an orthonormal set $\{|x_j\ra :1\le j\le k \}$ in $\IC^n$,
\begin{eqnarray*}
\hspace{-2cm}
\sum_{i,j=1}^k E_{ij} \otimes L_\gamma(|x_i\ra \la x_j|)
&=& \sum_{i,j=1}^k E_{ij} \otimes \(\gamma(\tr |x_i\ra \la x_j|)I_n - |x_i\ra\la x_j| \) \\
\hspace{-2cm}
&=& \sum_{i,j=1}^k E_{ij} \otimes \(\gamma(\tr \la x_j|x_i\ra)I_n - |x_i\ra \la x_j| \)
= \gamma I_{kn} - | x\ra\la x|.
\end{eqnarray*}
Since $|x\ra\la x|$ is a rank
one hermitian matrix with trace $k$, by Proposition \ref{2.1},
$L_\gamma$ is $k$-positive if and only if $\gamma \ge k$. \qed

Recall that a permutation $\pi$ of $(i_1, \dots, i_\ell)$ is   an
$\ell$-cycle if $\pi(i_j) = i_{j+1}$ for $j = 1, \dots, \ell-1$ and
$\pi(i_\ell) = i_1$. Note that every  permutation $\pi$ of
$(1,\dots, n)$ has a disjoint cycle decomposition
$\pi=(\pi_1)(\pi_2) \cdots (\pi_r)$, that is, there exists a set
$\{F_s\}_{s=1}^r$ of disjoint cycles of $\pi$ with $\cup_{s=1}^r
F_s=\{1,2,\ldots , n\}$ such that   $\pi_s=\pi|_{F_s}$ and
$\pi(i)=\pi_s(i)$ whenever $i\in F_s$. We have the following.

\begin{proposition} \label{6.2}
Suppose $\pi$ is a permutation of $(1,2,\ldots ,n)$ with disjoint
cycle decomposition $(\pi_1) \cdots  (\pi_r)$ such that the maximum
length of $\pi_i$ is equal to $\ell>1$ and
$P_{\pi}=\(\delta_{i\pi(j)}\)$ is the permutation matrix associated
with $\pi$. For $t\ge 0$, let $\Phi_{t,\pi}: M_n \rightarrow M_n$ be
the $D$-type map of the form (\ref{map}) with $D = (n-t)I_n +
tP_{\pi}$. Then $\Phi_{t,\pi}$ is positive if and only if $ t\leq
\frac{n}{\ell}$.
\end{proposition}

\it Proof. \rm  It is easily checked that for $0\le t\le 1$, the function
\begin{equation}\label{7}
\hspace{-1cm}
g(r_1,r_2,\ldots,r_s)=\sum_{i=1}^s\frac{1}{s-t+tr_i}\leq  1
\quad\hbox{for all $r_i>0$ and $r_1r_2\cdots r_s=1$,}
\end{equation}
and the  function $g$ attains the maximum $1$   when
$r_1=\cdots=r_s=1$.

Suppose $0\leq t\leq
\frac{n}{\ell}$. We are going to use condition  (2) in Corollary \ref{5.2} to show that $\Phi_{t,\pi}$ is positive.
For any   vector $|u\ra=(u_1, u_2, \ldots ,u_n)^t\in
{\IC}^{n}$, with $u_i\ne 0$ for all $i=1,\dots,n$, we have $f_i(u)=(n-t)|u_i|^2+t|u_{\pi(i)}|^2$. So, by
Corollary 5.2, $\Phi_{t,\pi}$ is positive if
\begin{equation} \label{fu} f(u_1,u_2,\ldots,u_n)
=\sum_{i=1}^n \frac{|u_i|^2}{(n-t)|u_i|^2+t|u_{\pi(i)}|^2}\leq
1\end{equation} for all vector $|u\ra=(u_1, u_2, \ldots ,u_n)^t$ with nonzero entries.

Suppose $\pi$ is a product of $r$ disjoint cycles, that is, $\pi=(\pi_1)(\pi_2)  \cdots  (\pi_r)$.
Let $F_j$ be the set of indices corresponding to the cycle $\pi_j$ and
$\ell_j$ denote the number of elements in $F_j$
for $j = 1,\dots,r$. Then
$\ell = \max\{\ell_1,\dots,\ell_r\}$ and $\sum_j \ell_j = n$.
For any  vector $|u\ra=(u_1, u_2, \ldots ,u_n)^t\in
{\IC}^{n}$,  with $u_i\ne 0$ for all $i=1,\dots, n$,
we have $\prod_{i\in F_j} \frac{|u_{\pi_j(i)}|^2}{|u_i|^2} = 1$.
It follows that
\begin{eqnarray*}
 f(u_1,u_2,\ldots,u_n)
&=&\sum_{i=1}^n\frac{|u_i|^2}{(n-t)|u_i|^2+t|u_{\pi(i)}|^2}\\
&=&\sum_{j=1}^r\sum_{i\in F_j}\frac{|u_i|^2}{(n-t)|u_i|^2+t|u_{\pi_j(i)}|^2}\\
&=&\sum_{j=1}^r\dfrac{\ell_j}{n} \, \sum_{i\in F_j}\frac{1}{ \ell_j-\frac{\ell_j}{n} t +\frac{\ell_j}{n}t \frac{|u_{\pi_j(i)}|^2}{|u_i|^2}}\\
&\leq &\sum_{j=1}^r\dfrac{\ell_j}n\, \cdot 1
=1,
\end{eqnarray*}
whenever $0 \le \frac{\ell_j}{n} t \le 1$ for all $1\le j \le r$ by (\ref{7}),
or equivalently, $0 \le t \le \frac{n}{\ell_j} \le \frac{n}{\ell}$.
Therefore, (\ref{fu}) holds.

\medskip
Conversely, suppose $t>\frac{n}\ell$. Let $\pi=(\pi_1)(\pi_2)\cdots
(\pi_r)$ be a decomposition of $\pi$ into disjoint cycles. Without
loss of generality, we may assume that $\ell_1=\ell\ge \ell_j$ for
all $j=2,\dots,r$, and $\pi_1$ is a cycle on $(1,2,\dots,\ell)$. Let
$u_i=\epsilon^{\frac{i}2}$, where $0<\epsilon< 1-\frac{n}{\ell t}$
for $i=1,\dots,\ell$ and $u_i=1$ for $\ell+1\le i\le n$. Then we
have
\begin{eqnarray*}
f(u_1,u_2,\ldots,u_n)
&=&\sum_{i=1}^{\ell-1}\frac{1}{(n-t) +t\epsilon}+\frac{1}{(n-t) +\frac{t}{\epsilon^{\ell-1}} }+\sum_{i=\ell+1}^n\frac{1}{(n-t) +t }\\
&\ge &\dfrac{\ell-1}{(n-t)+t\epsilon}+\dfrac{n-\ell}{n}\\
&>&\dfrac{\ell-1}{(n-t)+t\(1-\frac{n}{\ell t}\)}+\dfrac{n-\ell}{n}\\
&=&\dfrac{\ell-1}{ n- \frac{n}{\ell  } }+\dfrac{n-\ell}{n}
=\dfrac{\ell }{ n  }+\dfrac{n-\ell}{n}
= 1,
\end{eqnarray*}
which implies $\Phi_{t,\pi}$ is not positive.
  \qed

Next, we consider a general map $\Lambda_D$ of the form (\ref{map}).

\begin{proposition} \label{6.3} Let $\Lambda_{D}: M_n \rightarrow M_n$ have the form (\ref{map}) for a nonnegative matrix  $D = (d_{ij})$ with all row sum and column sum
equal to $n$. Then $\Lambda_{D}$ is positive if $d_{ii}\ge (n-1)$
for all $i = 1, \dots, n$. Moreover, the following conditions are equivalent.

\medskip\centerline{
{\rm (a)} \ $\Lambda_{D}$ is completely positive.
\qquad {\rm (b)} \ $\Lambda_{D}$ is 2-positive.
\qquad {\rm (c)} \ $D = nI_n$.
}
\end{proposition}

\it Proof. \rm Suppose $d_{ii} \ge n-1$ for all $i = 1, \dots, n$. Then $D = (n-1)I + S$
for a doubly stochastic matrix $S$, which is a convex combination of
permutation matrices (e.g., see \cite[Theorem 8.7.1, pp. 527]{Horn}). We
may represent $S$ as
$$S=\sum _{i=1}^m p_i P_{\pi_i}$$
for some permutations $\pi_1,\pi_2,\ldots,\pi_m$ of $\{1,2,\ldots, n\}$
and positive scalars $p_i$
with $\sum_{i=1}^m p_i=1$. Let
$S_i=(n-1)I_n+P_{\pi_i}$ and $\Lambda_{S_i}$ be the linear map of the
form as in (\ref{map}).  By Proposition \ref{6.2}, $\Lambda_{S_i}$ is a positive map.
Thus, $\Lambda_{D}$ is a
convex combination of positive maps, and is therefore positive.

\medskip
Next, we prove the three equivalent conditions. The implication.
(a) $\Rightarrow$ (b) is clear.
For (c) $\Rightarrow$ (a),
it is well known and easy to check, say, by considering the Choi matrix,
that $\Lambda_{D}$ is completely positive if $D = nI_n$.

It remains to prove (b) $\Rightarrow$ (c).
Suppose $D \ne n I_n$. Then $d_{ii} < n$ for some $i$. Without loss
of generality, we assume that $i = 1$.
Let 
$$U =\pmatrix{ |u_1\ra & \cdots & |u_n\ra} =   \frac{1}{\sqrt n} \pmatrix{1 & 0 & \cdots & 0 \cr 0 & 1 & \cdots & 1\cr} \in M_{2,n}.$$
Then
$$D_j := U \diag(d_{1j}, d_{2j}) U^\dag = \frac{1}{n}\pmatrix{d_{1j} & 0 \cr 0 & n-d_{1j}},
\quad j = 1, \dots, n.$$
As $n > d_{11}$,
$$
\la \hat u_1| D_1^{[-1]} |\hat u_1\ra = \frac{1}{d_{11}} > \frac{1}{n}\quad\hbox{and}\quad
\la\hat u_j| D_j^{[-1]} |\hat u_j\ra
= \frac{1}{n - d_{1j}} \ge \frac{1}{n} \quad \hbox{ for } j = 2, \dots,n.$$
Hence,
$$\sum_{j=1}^n \la \hat u_j| D_j^{[-1]} |\hat u_j\ra > 1,$$
and $\Lambda_D$ is not 2-positive by Proposition \ref{5.1}.
\qed

In \cite{QH}, the positive map $\Lambda_D$ with $D = (n-1)I_n + P$ for a permutation
matrix $P$ was considered, and the special case when $P$ is a length $n$-cycle
was discussed in details.
By Propositions \ref{6.2} and \ref{6.3}, we have the following corollary.

\begin{corollary} \label{6.4} Let
$\Lambda_D: M_n \rightarrow M_n$ be a $D$-type map of the form (\ref{map}) with
$D = (n-1)I_n + P$ for a  permutation matrix $P$.
Then $\Lambda_D$ is positive. Moreover, the following are equivalent.

\medskip
\centerline{
{\rm (a)} $\Lambda_D$ is completely positive. \quad {\rm (b)} $\Lambda_D$
is 2-positive. \quad {\rm (c)}  $D = nI_n$.}
\end{corollary}

The condition $d_{ii}\geq n-1$ for each $i$ is not necessary for
$\Lambda_D$ in Proposition \ref{6.3} to be positive as seen below.

\begin{example}\label{6.5} Let $D=\(\begin{array}{ccc}1.35& 1& 0.65\\&&\\  0.65& 1.35&1\\&&\\ 1& 0.65& 1.35\end{array}\)$. Here, $d_{ii}< 2=3-1$. Direct computation shows that
 $ \sum_{j=1}^3 \frac{|u_j|^2}{f_{j}(u)} \leq 1$ for all $(u_1,u_2,u_3) \in \IC^3$. Therefore, $\Lambda_D$ is positive by Corollary \ref{5.2}.
 \end{example}

\begin{example} \label{6.6} In Proposition \ref{6.2}, let $0\le t\le 1$ and $D=(d_{ij})=(n-t)I_n+tS$, where
 $S=\left(\begin{array}{cccc} s_1 & s_2 & \cdots &
s_n\\
s_n & s_1 & \cdots & s_{n-1} \\
\vdots & \vdots & \ddots &\vdots \\
s_2 &s_3 &\cdots &s_1 \end{array}\right)$ with $s_i\geq 0$
($i=1,2,\ldots, n$) and $\sum_{i=1}^n s_i=1$. Define $\Lambda_D: M_n \to M_n$ by
$$\Lambda_D((a_{ij}))=\left(\begin{array}{cccc} f_1 & -a_{12} & \cdots &
-a_{1n}\\
-a_{21} & f_2 & \cdots & -a_{2n} \\
\vdots & \vdots & \ddots &\vdots \\
-a_{n1} & -a_{n2} &\cdots &f_n \end{array}\right),$$
where
\begin{eqnarray*}
f_1 &=&(n-t-1+ts_1)a_{11}+ts_na_{22}+ts_{n-1}a_{33}+\cdots+ts_2a_{nn}, \\
f_2 &=& ts_2a_{11}+(n-t-1+ts_1)a_{22}+ts_na_{33}+\cdots+ts_{3}a_{nn}, \\
&\vdots&  \\
f_n &=& ts_na_{11}+ts_{n-1}a_{22}+ts_{n-2}a_{33}+\cdots+(n-t-1+ts_{1})a_{nn}.
\end{eqnarray*}
By Proposition \ref{6.2}, the map $\Lambda_D$ is positive.
\end{example}

Finally, we give an example which illustrates how to apply
Proposition \ref{6.2} to construct positive elementary operators for any dimension.

\begin{example} \label{6.7} Let $H$ and $K$ be  Hilbert spaces of dimension at least $n$,
and let $\{|e_i\rangle\}_{i=1}^n$ and
$\{|\hat e_j\rangle\}_{j=1}^n$ be any orthonormal sets of $H$ and $K$,
respectively. For any permutation $\pi\not={\rm id}$ of
$\{1,2,\ldots, n\}$, let $l(\pi)=l\leq n$.
Let
$\Phi_{t,\pi}:{\mathcal B}(H)\rightarrow {\mathcal B}(K)$ be defined
by
$$
\Phi_{t,\pi}(A)=(n-t)\sum_{i=1}^nE_{ii}AE_{ii}^\dagger+
t\sum_{i=1}^{n}E_{i,\pi(i)}AE_{i,\pi(i)}^\dagger
-(\sum_{i=1}^nE_{ii})A(\sum_{i=1}^nE_{ii})^\dagger
\hbox{ for all $A\in{\mathcal B}(H)$},$$
where  $E_{ji}=|\hat e_j\rangle\langle
e_i|$. Then $\Phi_{t,\pi}$ is positive if and only if $0\leq t\leq
\frac{n}{l}$.
\end{example}

In fact, for the case $\dim H=\dim K=n$, $\Phi_{t,\pi}$ is a
$D$-type  map of the form (\ref{map}) with $D=(n-t)I+tP_\pi$
as discussed in Proposition  \ref{6.2}.

\section{Decomposable $D$-type positive maps}

Decomposability of positive linear maps is a topic of particular
importance in quantum information theory
since it is related to the PPT states (that is, the
states with positive partial transpose). In this section, we will
give a new class of decomposable positive linear maps.

The following result is well known (see \cite{HH}).

\begin{proposition}\label{7.1} Suppose $L: M_n\rightarrow M_m$ has the form (\ref{map}).
Then $L$ is decomposable if and only if the Choi matrix $C(L)$
is a sum of two matrices $C_1$ and $C_2$ such that
$C_1$ and the partial transpose of $C_2$ are positive semi-definite.
\end{proposition}

In \cite{QH}, it was shown that the linear maps
$\Phi^{(k)}=\Phi_{1,\pi}$ with $\pi(i)=i+k$ (mod $n$) in
Proposition \ref{6.2} are indecomposable whenever either $n$ is odd or
$k\not=\frac{n}{2}$. It was asked in \cite{QH} that whether or not
$\Phi^{(\frac{n}{2})}$ is decomposable when $n$ is even. In this
section, we will answer this question by showing that
$\Phi^{(\frac{n}{2})}$ is decomposable. In fact, this is a special
case of the following proposition as $(\pi)^2={\rm id}$.

\begin{proposition}
Let $\pi$ be a permutation of $\{1,2,\ldots, n\}$. If $\pi^2={\rm
id}$, then the positive linear map $\Phi_{1,\pi}$ in Proposition \ref{6.2} is
decomposable.
\end{proposition}

\it Proof. \rm For simplicity, denote $\Phi = \Phi_{1,\pi}$.
Let $F$ be the set of fixed points of $\pi$. Since
$\Phi(E_{ii})=(n-2)E_{ii}+E_{\pi(i),\pi(i)}$ and
$\Phi(E_{ij})=-E_{ij}$, the Choi matrix of $\Phi$ is
\begin{eqnarray*}
C(\Phi) &=&\sum_{i=1}^n(n-2)E_{ii}\otimes E_{ii}+ \sum_{i=1}^nE_{\pi(i),\pi(i)}\otimes E_{ii}-\sum_{i\not=j}E_{ij}\otimes E_{ij} \\
&=& \sum_{i\in F}(n-1)E_{ii}\otimes E_{ii}+\sum_{i\not\in
F}(n-2)E_{ii}\otimes E_{ii} \\
&& -\sum_{i\neq j; \pi(i)\neq j}E_{ij}\otimes E_{ij}+\sum_{i\not\in
F}E_{\pi(i),\pi(i)}\otimes E_{ii}-\sum_{i\not\in
F}E_{i,\pi(i)}\otimes E_{i,\pi(i)}.
\end{eqnarray*}
Let
$$C_1=\sum_{i\in F}(n-1)E_{ii}\otimes E_{ii}+\sum_{i\not\in
F}(n-2)E_{ii}\otimes E_{ii}-\sum_{i\neq j; \pi(i)\neq
j}E_{ij}\otimes E_{ij}$$and
$$C_2=\sum_{i\not\in
F}E_{\pi(i),\pi(i)}\otimes E_{ii}-\sum_{i\not\in
F}E_{i,\pi(i)}\otimes E_{i,\pi(i)}.$$ Since $\pi^2={\rm id}$, the
cardinal number of $F^c$ must be even. Thus we have
$$C_2=\sum_{i< \pi(i)}(E_{\pi(i),\pi(i)}\otimes E_{ii}+E_{ii}\otimes E_{\pi(i),\pi(i)}-E_{i,\pi(i)}\otimes E_{i,\pi(i)}-E_{\pi(i),i}\otimes E_{\pi(i),i}).$$
As
$$C_2^{{\rm T}_2}=\sum_{i< \pi(i)}(E_{\pi(i),\pi(i)}\otimes E_{ii}+E_{ii}\otimes E_{\pi(i),\pi(i)}
-E_{i,\pi(i)}\otimes E_{\pi(i),i}-E_{\pi(i),i}\otimes E_{i,\pi(i)})\geq 0,$$
we see that $C_2$ is PPT.

Observe that $C_1\cong A\oplus 0$, where $A=(a_{ij})\in M_n$ is a
Hermitian matrix satisfying $a_{ii}=n-2$ or $n-1$, $a_{ij}=0$ or
$-1$ so that $\sum_{j=1}^n a_{ij}=0$. It is easily seen from the
strictly diagonal dominance theorem  (Ref. \cite[Theorem 6.1.10, pp.
349]{Horn}) that $A$ is positive semi-definite. So $C_1\geq 0$, and by
Proposition 7.1, $\Phi$ is decomposable. \qed

\section{Conclusion}

Because $k$-positive linear maps are important in theory as well as applications, many
researchers have been working on problems such as finding efficient criteria to determine
$k$-positive maps and constructing $k$-positive maps with simple structure. In this paper,
we present some existing and new criteria for $k$-positive maps. Using these criteria, we
are able to improve the results of other researchers.  Moreover, new classes of $k$-positive
maps are introduced, and the decomposability of the maps are discussed. These lead
the answers of some open problems.

There are other interesting topics on positive maps relative to
information science such as indecomposability, atomic, optimality,
etc.. These questions for $D$-type positive maps constructed from a
permutation as in Proposition 6.2 are studied by Hou and Qi in other
papers.


\section*{Acknowledgments}

Research of Hou was supported by the NNSF of China (11171249) and a
grant from International Cooperation Program in Sciences and
Technology  of Shanxi (2011081039). Research of Li was supported by
the 2011 Shanxi 100 Talent program, a USA NSF grant, and a HK RGC
grant. He is an honorary professor of University of Hong Kong  and
Shanghai University. Research of Poon was supported by a USA NSF
grant and a HK RGC grant. Research of Qi was supported by the NNSF
of China (11101250) and Youth Foundation of Shanxi Province
(2012021004). Research of Sze was supported by a HK RGC grant PolyU
502411. 

\end{document}